\documentclass[11pt]{cernrep}
\usepackage{graphicx}
\usepackage{here}
\begin{document}
 \title{Support Vector Machines in Analysis of Top Quark Production}
\author{A. Vaiciulis}
\institute{University of Rochester, Rochester, New York, USA}
\maketitle
\begin{abstract}
Multivariate data analysis techniques have the potential
to improve physics analyses in many ways. The common
classification problem of signal/background discrimination
is one example. A comparison of a conventional method and 
a Support Vector Machine algorithm is presented here for 
the case of identifying top quark signal events in the 
dilepton decay channel amidst a large number of background 
events.
\end{abstract}

\section{INTRODUCTION}

A common problem in high energy physics is that of classification.
Is it a signal or background event? Does the energy deposit correspond
to a tau particle or not? There are many problems for which no
explicit method exists to determine the correct output from the input data.
One approach is to devise an algorithm which finds and exploits complex patterns
in input/output examples (labeled training data) to learn the
solution to the problem. This is called ``supervised learning''.
Such an algorithm can map each training example onto two categories
(``binary classification''), more than two categories (``multi-class 
classification''), or continuous, real-valued output (``regression'').

One possible goal is to modify the learning algorithm in an iterative
procedure until all training data are classified correctly (i.e.
no mistakes). A potential problem with this goal is noisy training
data. There may be no correct underlying classification function. Two 
similar training examples may be in different categories. An example
is distinguishing a photon deposit in a calorimeter from a 
$\pi^o\rightarrow\gamma\gamma$ deposit. Another problem may be that the
resulting algorithm misclassifies unseen data because it has ``overfit''
the training data.
A better goal is to optimize ``generalization'' -- the ability to correctly
classify unseen data. In this approach we should not add extra complexity
unless it causes a significant improvement in performance.
Potential problems include
making mistakes due to local minima, overfitting when using a complex 
mapping function on a small training set, and having a large number
of tunable parameters which makes the algorithm difficult to use.

Section 2 of this paper shows how the Support Vector Machine learning
methodology addresses these problems. The description closely follows that
of several references in the literature \cite{crist,burges,schoel}. 
The use of hyperplane
classifiers in SVMs is described first for the linear, separable case.
This is followed by the extension to nonlinear SVMs and to nonseparable 
data. In section 3, the results of a study using SVMs in
an analysis of top quark production are presented.

\section{SUPPORT VECTOR MACHINE METHODOLOGY}

\subsection{Overview}

In the early 1960s the support vector method was developed
to construct separating hyperplanes for pattern recognition problems 
\cite{vapnik63,vapnik64}. In the 1990s
it was generalized for constructing nonlinear separating functions 
\cite{boser,cortes} and for estimating real-valued functions (regression)
\cite{vapnik95}. 
Current activities \cite{kernelmach} include a special SVM issue of the journal 
Neurocomputing (2002), 
a NATO Advanced Study Institute on Learning Theory and Practice (July, 2002),
an International Workshop on Practical Application of Support Vector Machines 
in Pattern Recognition (August, 2002), and a Special Session on Support 
Vector Machines at the International Conference on Neural 
Information Processing (November, 2002).

Applications of SVMs include text categorization, character recognition, 
bioinformatics and face detection. The main idea of the SVM approach
is to map the training data into a high 
dimensional feature space in which a decision boundary is determined
by constructing the optimal separating hyperplane. Computations in
the feature space are avoided by using a kernel function. This approach
uses concepts from statistical 
learning theory to describe which factors have to be controlled for
good generalization.

\subsection{Generalization and Capacity}

The formal goal is to estimate the function $f:\Re^N \rightarrow \{\pm1\}$
using input/output training data
\begin{center}
$(\vec{x}_1,y_1),...,(\vec{x}_\ell,y_\ell) \in \Re^N\times\{\pm1\}$
\end{center}
such that $f$ will correctly classify unseen examples $(\vec{x},y)$,
i.e. $f(\vec{x}) = y$. $\ell$ is the number of training examples.
There is a tension between the function's complexity and the resulting 
accuracy.
According to statistical learning theory, for good generalization
we should restrict the class of functions from which $f$ is chosen.
Simply minimizing the training error, 
\begin{center}
$(1/\ell)\sum_{i}\mid f(\vec{x}_i)-y_i \mid $, 
\end{center}
does not necessarily result in good generalization. More precisely, 
we restrict the class
of functions to one with a ``capacity'' suitable for the amount of available
training data. 
The ``capacity'' is the richness or flexibility of the function class. 
Low capacity leads to good generalization,
regardless of the dimensionality of the space, assuming the function describes 
the data well.
Controlling the capacity is one way to improve generalization accuracy.

\subsection{Hyperplane Classifiers}

Support Vector classifiers are based on the class of hyperplanes
\begin{center}
$(\vec{w}\cdot\vec{x})+b = 0$
\end{center}
with $\vec{w} \in \Re^N,$ $b \in \Re$
and corresponding to the decision function
\begin{center}
$f(\vec{x}) =$ sign$[(\vec{w}\cdot\vec{x}) + b]$.
\end{center}
$\vec{w}$ is called the ``weight vector'' and b the ``threshold''.
$\vec{w}$ and b are the parameters controlling the function and must 
be learned from the data.
For pedagogical purposes we are considering first the 
linear, separable, binary classification case, i.e. $f(\vec{x})$
is a linear function of $\vec{x}$ and there are two classes which can
be separated completely.

\begin{figure}
\begin{center}

\hspace{-1.5 in}
\begin{minipage}[t]{1.5in}
\includegraphics[width=5cm]{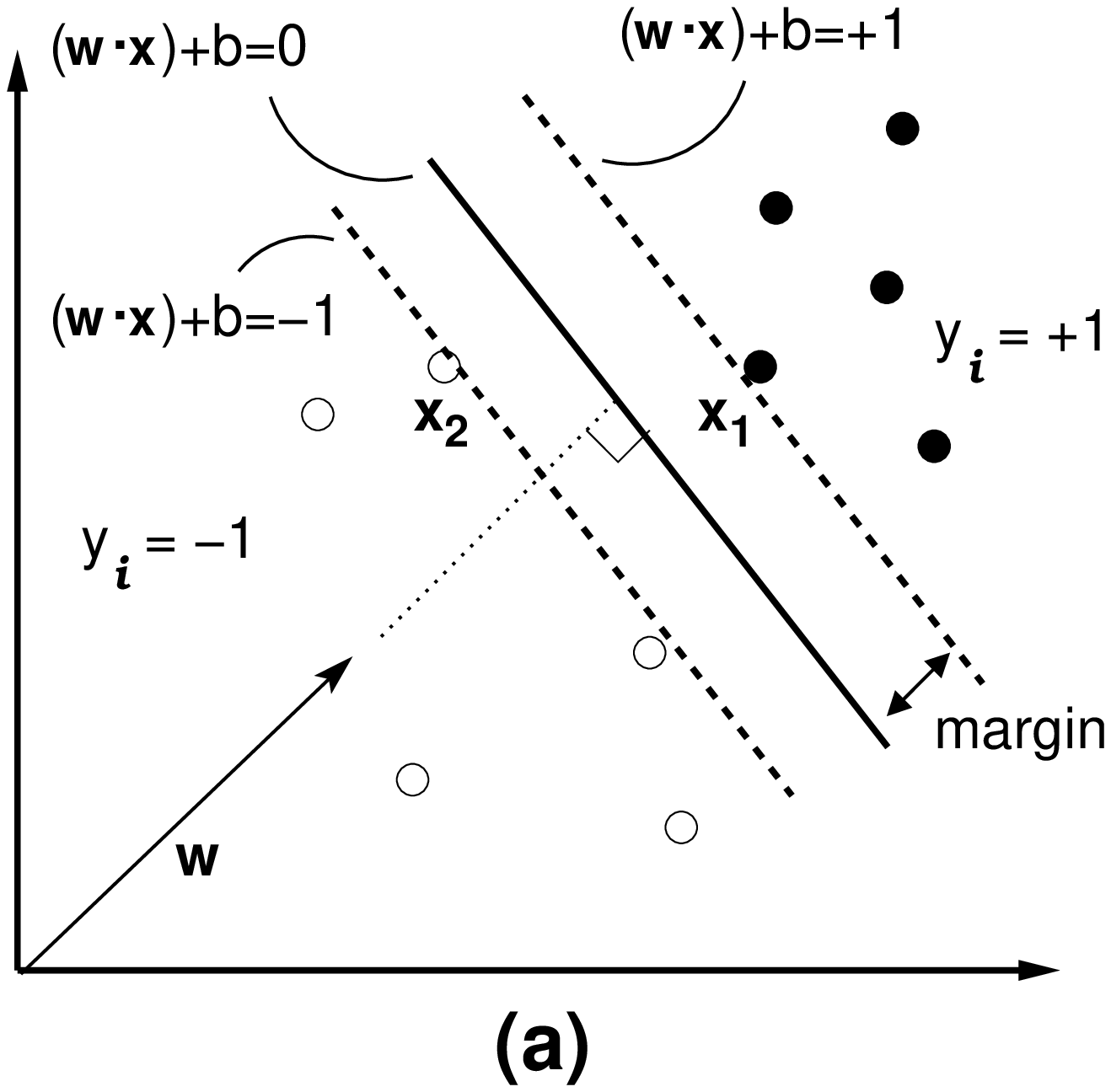}
\end{minipage}
\hspace{1.0 in}
\begin{minipage}[t]{1.5in}
\includegraphics[width=7cm]{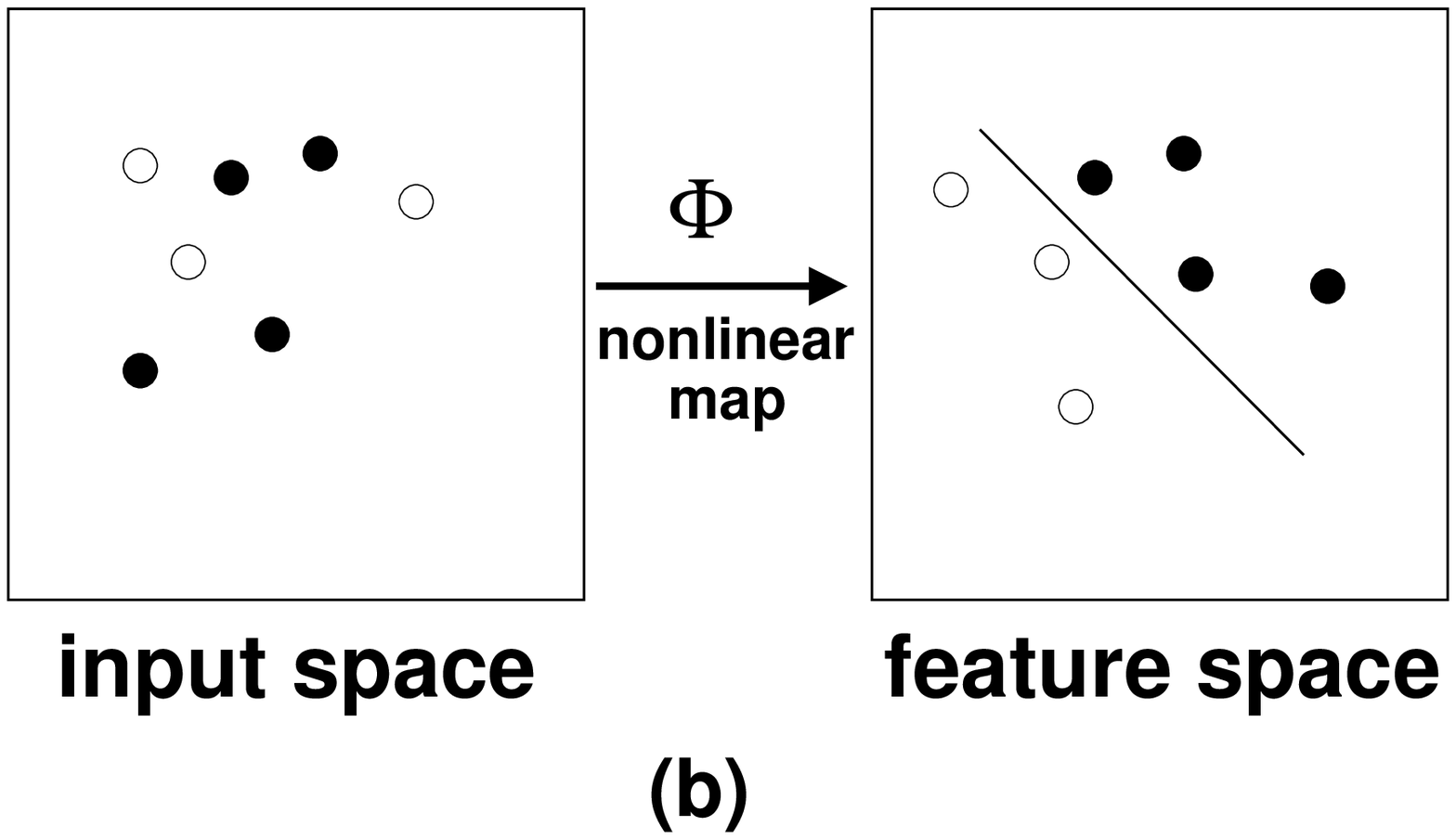}
\end{minipage}

\caption{(a) Geometric interpretation of hyperplane classifier in two dimensions. 
(b) Cartoon showing how a nonlinear problem in input space is mapped onto a
linear problem in feature space.}
\label{fig1label}
\end{center}
\end{figure}

The unique hyperplane with maximal margin of separation between the two classes
is called the optimal hyperplane. It can be shown that it has the lowest
capacity of any hyperplane, which minimizes the risk of overfitting.
The optimization problem thus becomes one of finding the optimal hyperplane. 
This is different than the ``intuitive'' way of decreasing capacity by 
reducing the number of degrees of freedom (e.g. decreasing the number of
nodes or layers in a neural network).
A geometric interpretation is that the hyperplane splits the input space 
into two parts, each one corresponding to a different class. Figure 
\ref{fig1label}a
shows a two dimensional example with two classes denoted by solid 
circles ($y_i = +1$) and open circles ($y_i = -1$). The optimal hyperplane
is shown by the solid line between the two classes.

The size of the margin is inversely proportional to the norm of $\vec{w}$.
To find the optimal hyperplane, 
$\|\vec{w}\|^2$ must be minimized subject to constraints 
$y_i[ (\vec{w}\cdot\vec{x}_i) + b] \geq 1 $ for
$i = 1,...\ell$.
The $\vec{x}_i$ for which the equality holds are called ``support vectors''.
They carry all information about the problem. 
They lie on a hyperplane defining the margin and their removal would change
the solution. In Figure \ref{fig1label}a $\vec{x}_1$ and $\vec{x}_2$ 
are examples of support vectors.

To solve this constrained quadratic optimization problem, we first reformulate
it in terms of a Lagrangian,
\begin{center}
$\mathcal{L}(\vec{w}, b, \alpha_i ) = (1/2)\|\vec{w}\|^2 -
\sum_{i}\alpha_i[y_i((\vec{x}_i\cdot\vec{w}) + b) - 1] $.
\end{center}
This reformulation is done because it is easier to handle constraints 
on the Lagrange multipliers, $\alpha_i$ ($\alpha_i \geq 0$), and 
the training data will only appear in the form of dot products between 
vectors. This will allow us to generalize to the nonlinear case.
Note that the number of free parameters in an SVM increases as the number of
training examples increases.

Generalization theory indicates
how to control the capacity by controlling the margin of separation.
Specifically, we need to find the optimal hyperplane.
Optimization theory provides mathematical tools to find this hyperplane.
We need to minimize $\mathcal{L}$ with respect to $\vec{w}$ and b 
(\textit{primal variables}) and 
maximize $\mathcal{L}$ with respect to $\alpha_i$ (\textit{dual variables}).
The solution has an expansion in terms of a subset of input vectors 
(with $\alpha_i \neq 0$) called Support Vectors,
\begin{center}
$\vec{w} = \sum_{i}\alpha_iy_i\vec{x}_i$.
\end{center}
$\alpha_i$ corresponds to the difficulty in classifying 
the point. Small $\alpha_i$ means
easy classification.
In \textit{dual} form the optimization problem becomes one of finding
the $\alpha_i$ which maximize
\begin{center}
$\mathcal{L} = \sum_{i}\alpha_i - (1/2)\sum_{i,j}\alpha_i\alpha_jy_iy_j
\vec{x}_i\cdot\vec{x}_j$
\end{center}
subject to the constraints $\alpha_i \geq 0$ and $\sum_{i}\alpha_iy_i = 0$.
The decision function is 
\begin{center}
$f(\vec{x}) =$ sign$ [\sum_iy_i\alpha_i(\vec{x}\cdot\vec{x}_i) + b] $
\end{center}
Both the optimization problem and the final decision function
depend only on dot products between input vectors. This is crucial for the
successful generalization to the nonlinear case.

\subsection{Feature Spaces and Kernels}

If $f(\vec{x})$ is a nonlinear function of $\vec{x}$
one possible approach is to use a neural network, which consists of a 
network of simple linear classifiers. 
Problems with this approach include many parameters and the existence of
local minima.
The SVM approach is to map the input data into a high, possibly infinite
dimensional feature space,
$\mathcal{F}$, via a nonlinear map $\Phi:R^N\rightarrow\mathcal{F}$.
Then the optimal hyperplane algorithm can be used in $\mathcal{F}$ (see Figure
\ref{fig1label}b).
This high dimensionality may lead to a practical computational problem
in feature space. 
Since the input vectors appear in the problem only inside dot products, however,
we only need to use dot products in feature space. If we can find a 
kernel function, $\mathcal{K}$, such that
\begin{center}
$\mathcal{K}(\vec{x}_1,\vec{x}_2) = \Phi(\vec{x}_1)\cdot
\Phi(\vec{x}_2)$
\end{center}
then we don't need to know $\Phi$ explicitly.
Mercer's Theorem tells us that 
a function $\mathcal{K}(\vec{x},\vec{y})$ is a kernel, i.e. there
exists a mapping $\Phi$ such that 
\begin{displaymath}
\mathcal{K}(\vec{x}_1,\vec{x}_2) = \Phi(\vec{x}_1)\cdot\Phi(\vec{x}_2)
\end{displaymath}
if 
\begin{displaymath}
\int \mathcal{K}(\vec{x},\vec{y})g(\vec{x})g(\vec{y})d\vec{x}
d\vec{y} \geq 0
\end{displaymath}
for all g such that $\displaystyle \int g(\vec{x})^2d\vec{x}$ is finite.
Mercer's Theorem does not tell us how to construct $\Phi$ but this explicit
mapping is not needed to solve the problem. Rather than creating a function
and testing whether it is a kernel function, we can choose from known
kernel functions:
\begin{itemize}
  \item[--] $\mathcal{K}(\vec{x},\vec{y}) = (\vec{x}\cdot\vec{y})^d$\\
        (polynomial of degree d)
  \item[--] $\mathcal{K}(\vec{x},\vec{y}) = $
        exp$(-\parallel{\vec{x} - \vec{y}}\parallel^2/(2\sigma^2))$\\
        (Gaussian Radial Basis Function)
  \item[--] $\mathcal{K}(\vec{x},\vec{y}) =$ tanh$(\kappa
        (\vec{x}\cdot\vec{y}) + \theta)$\\
        (sigmoid)
\end{itemize}
Different kernel functions lead to similar classification accuracies 
and Support Vector sets.

\subsection{Nonlinear SVMs}

To extend the methodology described above to nonlinear problems, we
substitute $\Phi(\vec{x}_i)$ for each training example 
$\vec{x}_i$ and substitute the kernel for dot products of $\Phi$. The
decision function then becomes
\begin{center}
$f(\vec{x}) =$ sign$ [\sum_iy_i\alpha_i \mathcal{K}(\vec{x},\vec{x}_i) + b] $
\end{center}
and the optimization problem is one of maximizing
\begin{center}
$\mathcal{L} = \sum_{i}\alpha_i - (1/2)\sum_{i,j}\alpha_i\alpha_jy_iy_j
\mathcal{K}(\vec{x}_i,\vec{x}_j)$
\end{center}
subject to constraints $\alpha_i \geq 0$ and $\sum_{i}\alpha_iy_i = 0$.
Due to Mercer's conditions on the kernel, the corresponding
optimization problem is a well defined convex quadratic programming
problem which means there is a global minimum. This is an advantage
of SVMs compared to neural networks, which may only find a local
minimum.

\subsection{Nonseparable Data}

Section 2.3 described the SVM approach for linear, separable problems.
Sections 2.4 and 2.5 described the extension to nonlinear, separable
problems.
Real world applications, however, tend to have a large overlap of the 
two classes, i.e. nonseparable data. In general for these problems, 
a linear separation in feature space is not possible 
unless a very complex kernel is used which may lead to overfitting.
The  Lagrangian will grow arbitrarily large and 
the optimization problem will not converge.
So we introduce slack variables ($\zeta_i \geq 0$) to
allow for the possibility of points violating the constraints. Recall that 
in the separable case, there is no training error.
\begin{figure}
\begin{center}

\hspace{-1.5 in}
\begin{minipage}[t]{1.5in}
\includegraphics[width=5cm]{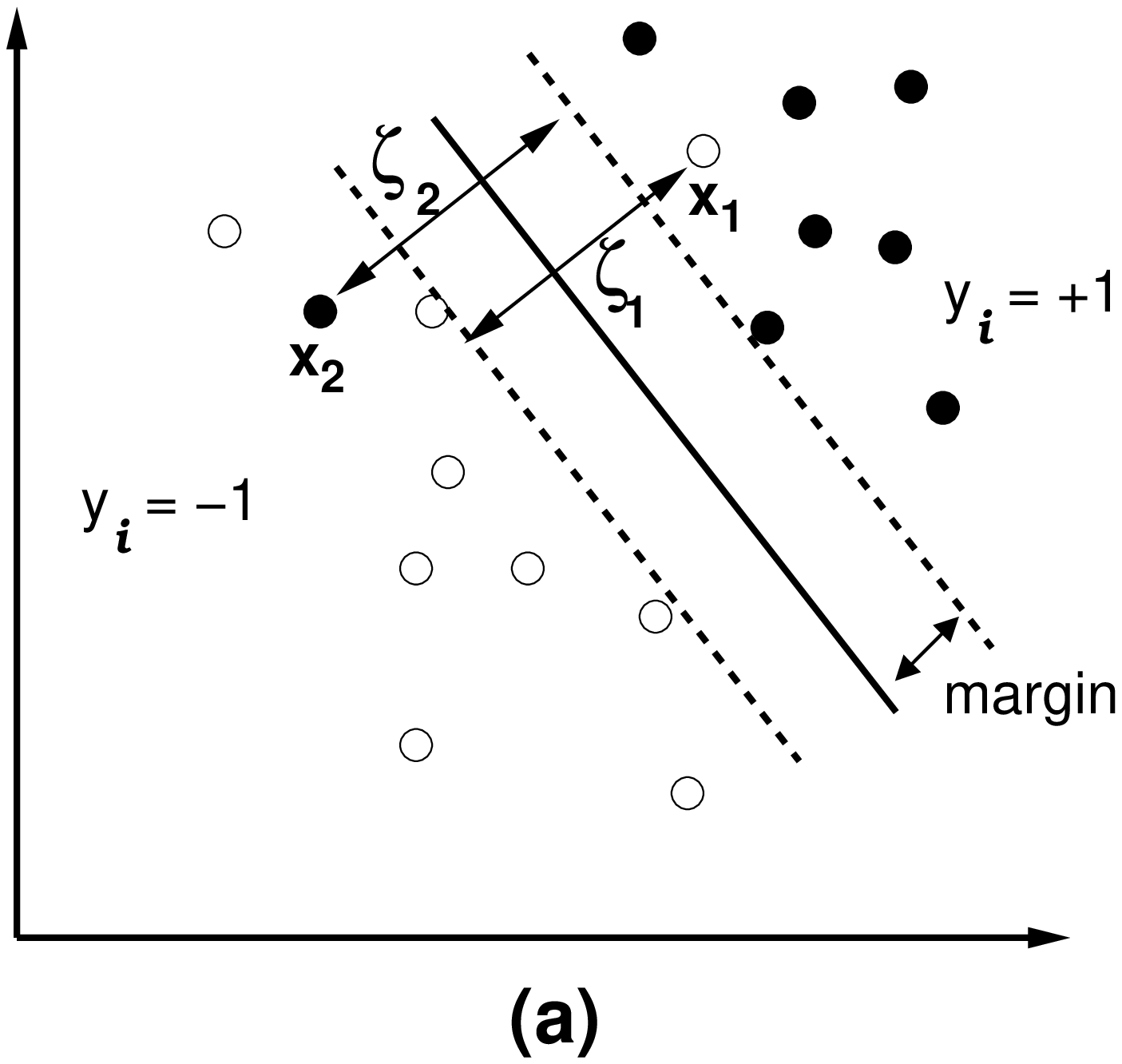}
\end{minipage}
\hspace{1.0 in}
\begin{minipage}[t]{1.5in}
\includegraphics[width=6cm]{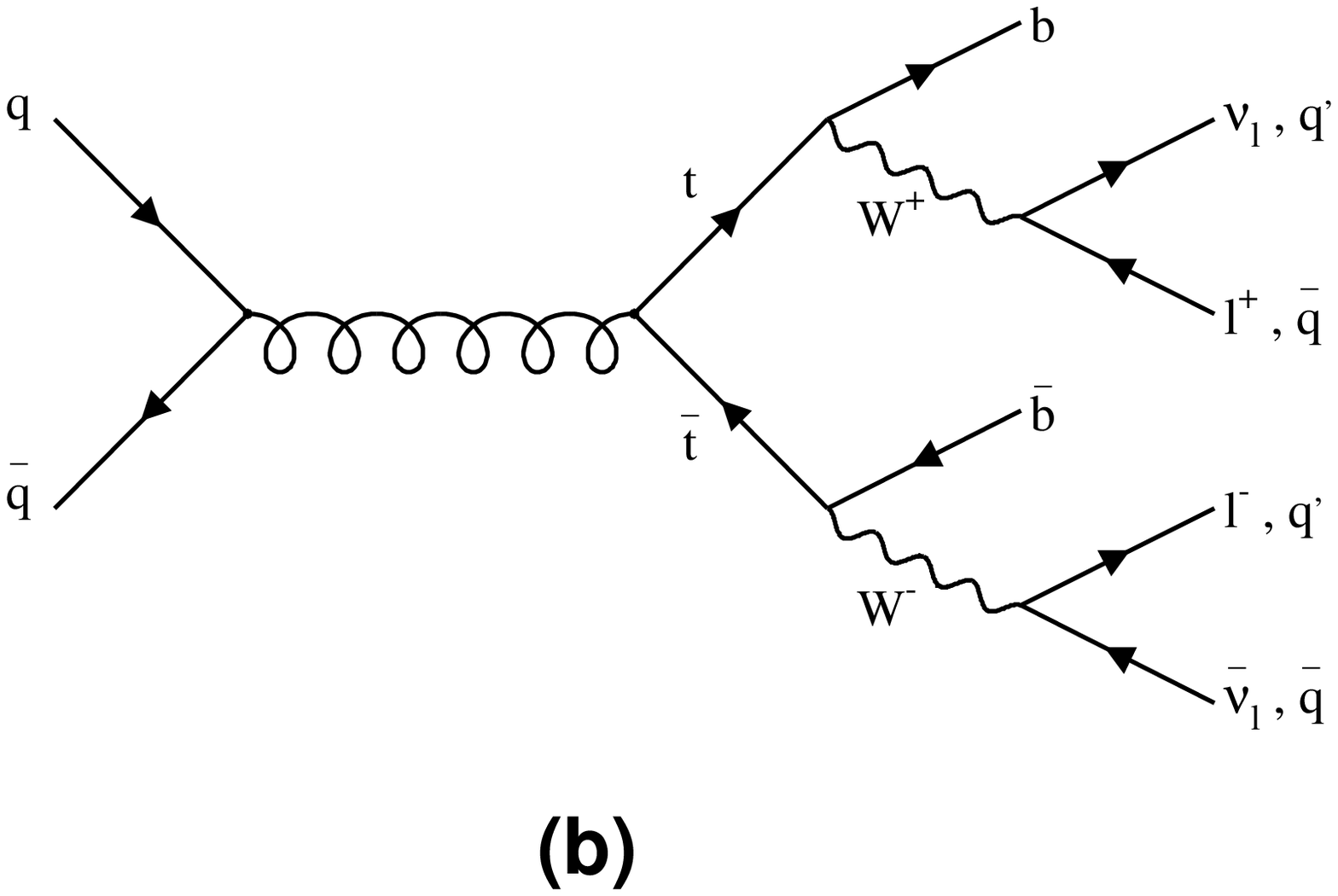}
\end{minipage}

\caption{(a) Two dimensional example of nonseparable data in which two
points are misclassified. (b) $t\bar{t}$ event topologies.}
\label{fig2label}
\end{center}
\end{figure}
The new, relaxed contraints are 
\begin{center}
$y_i( (\vec{w}\cdot\vec{x}_i) + b) \geq 1- \zeta_i $.
\end{center}
$\zeta_i > 0$ means $\vec{x}_i$ is misclassified. Figure \ref{fig2label}a 
shows an example in which the points $\vec{x}_1$ and $\vec{x}_2$
are on the wrong side of the decision boundary.
The slack variables allow for some number of errors in the optimization problem.
Good generalization is now achieved by controlling two things -- 
the capacity, as before (i.e. control margin size via $\|\vec{w}\|$)
and the number of training errors.
We can redo the Lagrangian formulation, adding a term 
C$\sum_i\zeta_i$ where
parameter C controls the error penalty. A large C causes a large penalty.
C is the only user chosen parameter aside from the kernel parameters.
This leads to the same dual optimization problem and constraints 
as for the separable case except 
\begin{center}
$\alpha_i \geq 0 \; \rightarrow \; 0\leq\alpha_i\leq C$.
\end{center}
This completes the brief description of the SVM approach to nonlinear,
nonseparable, binary classification problems. The extension to
regression is not included here but can be found in the literature
\cite{crist,burges,vapnik95}.

\section{SVM USE IN TOP QUARK ANALYSIS}

One possible application of SVMs in HEP is improving 
signal vs. background discrimination in the $t\bar{t}$ dilepton channel.
All direct measurements of the top quark are from the Fermilab $p\bar{p}$ 
collider Run 1 \cite{cdfd0findtop}.
According to the Standard Model,
the top quark is produced at the Tevatron mainly via $t\bar{t}$ pair production. 
The two main
processes are the $q\bar{q}$ annihilation diagram ($q\bar{q} \rightarrow t\bar{t}$) and gluon-gluon
fusion ($gg \rightarrow t\bar{t}$). These contribute about 90\% and 10\% 
respectively to the
$t\bar{t}$ cross section at the Tevatron, which has a Standard Model predicted
value of 4.7-5.5 pb \cite{berger}.
The expected $t\bar{t}$ cross section of $\sim$5.0 pb is a small fraction
of the total cross section; from
more than 10$^{12}$ $p\bar{p}$ collisions in Run 1, the top measurements are based
on $\sim$100 events.

In order to identify events in which top quarks are produced, the decay
products must be detected. According to the Standard
Model, the top quark decays to Wb with a branching ratio of nearly 100\% 
(Figure \ref{fig2label}b). The individual branching ratios for W decay are 
BR(W$^{+} \rightarrow e^{+}\nu$) = 1/9,
BR(W$^{+} \rightarrow \mu^{+}\nu$) = 1/9,
BR(W$^{+} \rightarrow \tau^{+}\nu$) = 1/9,
BR(W$^{+} \rightarrow q\bar{q}$) = 6/9.
This leads to four main $t\bar{t}$ event topologies:
dilepton (5\%), lepton +jet (30\%), all-hadronic (44\%) and events with taus 
(21\%).
The all-hadronic channel has large QCD backgrounds
($p\bar{p} \rightarrow$ six jets).
The dilepton channel is the most pure but has the fewest number of events because
of the low branching ratio. In this study we try to increase the signal
efficiency in the $e\mu$ dilepton channel.
Backgrounds include WW production and Z$\rightarrow\tau^{+}\tau^{-}$.
We consider only WW in this study.

\begin{figure}
\begin{center}
\hspace{-1.5 in}
\begin{minipage}[t]{1.5in}
\includegraphics[width=7.8cm]{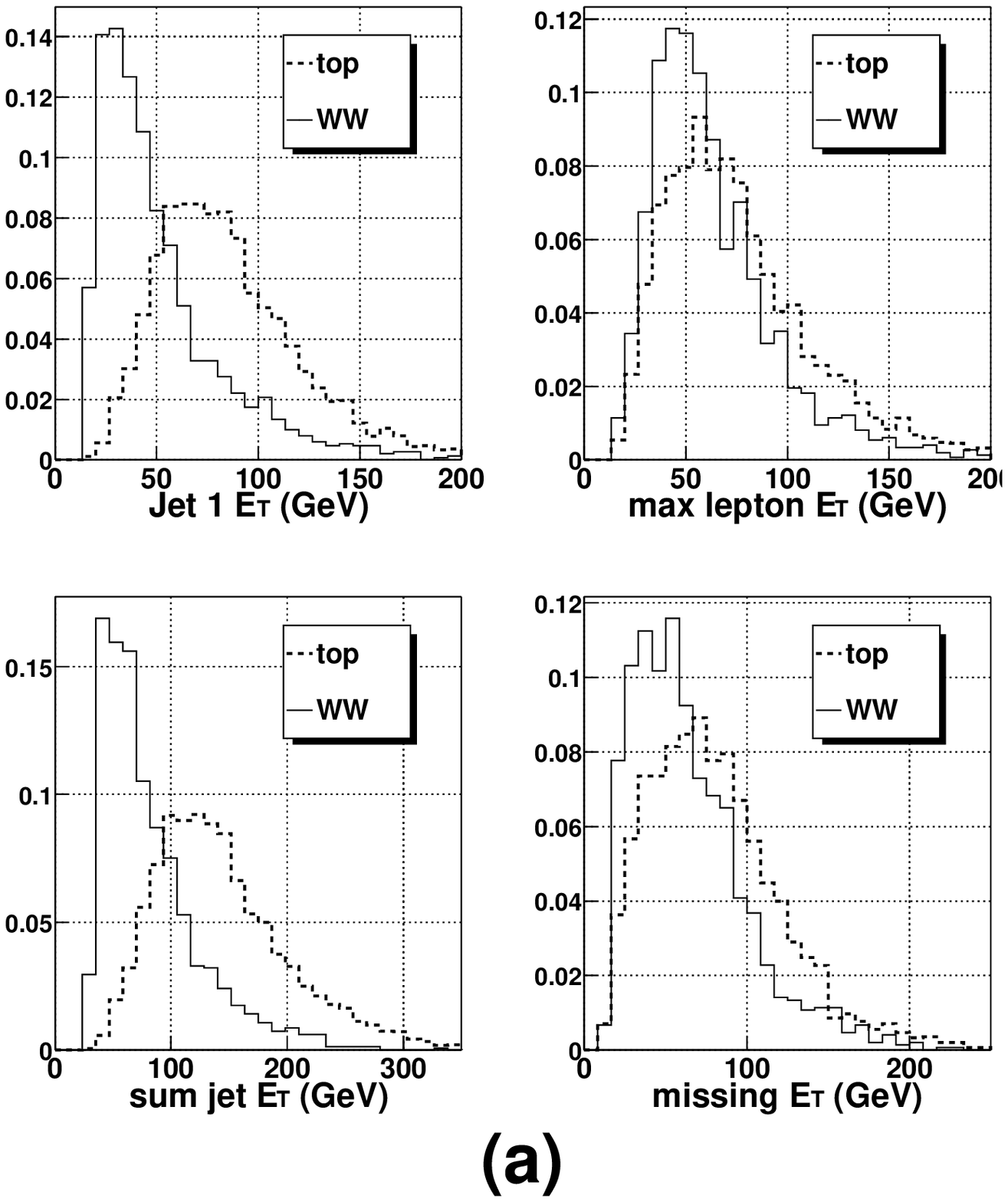}
\end{minipage}
\hspace{1.8 in}
\begin{minipage}[t]{1.5in}
\includegraphics[width=7.6cm]{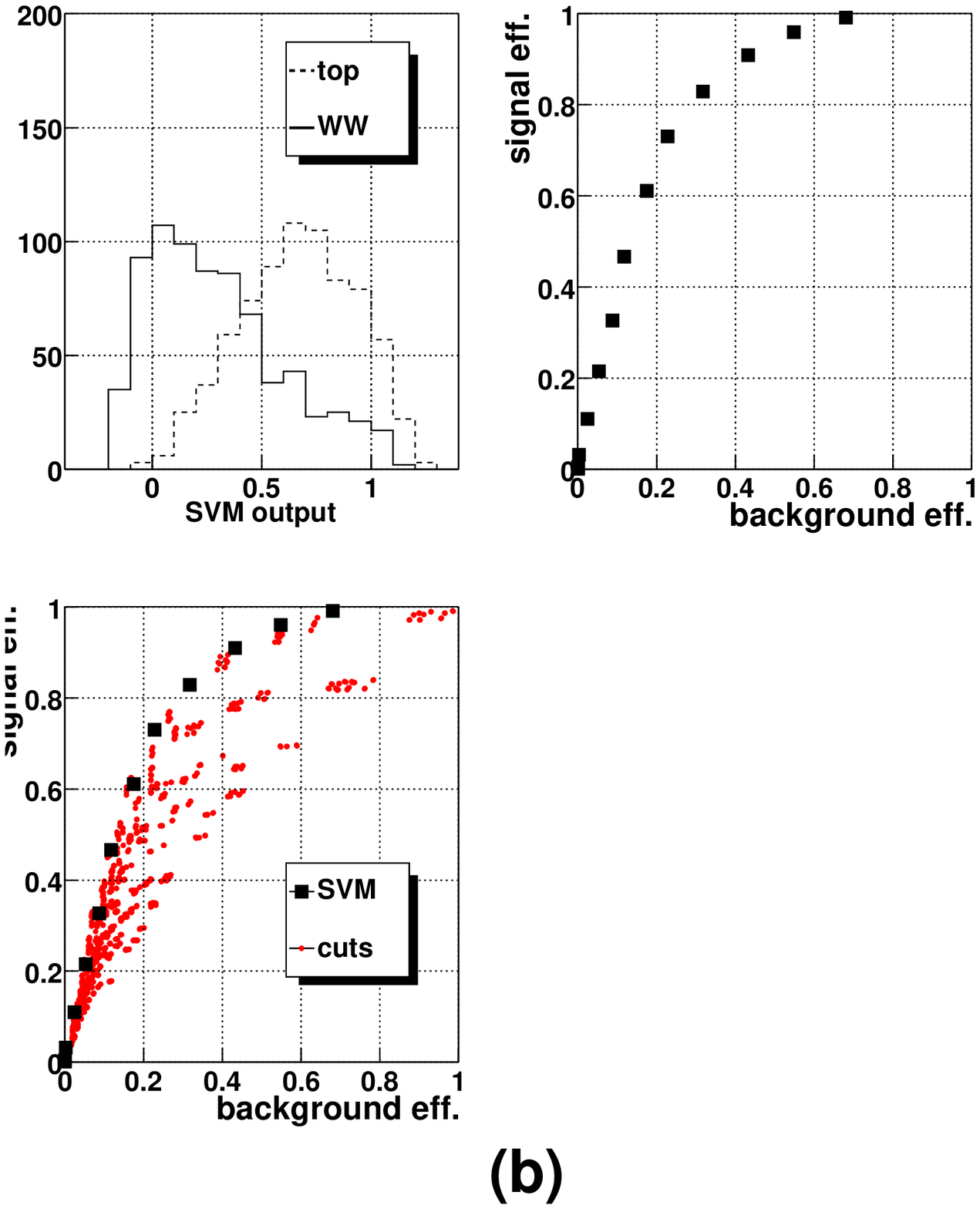}
\end{minipage}
\caption{(a) A comparison of the $t\bar{t}$ and WW distributions for the four 
input variables. (b) Signal efficiency versus background efficiency for
SVM compared with conventional cuts.}
\label{fig3label}
\end{center}
\end{figure}

The signature of an $e\mu$ dilepton event is 
two high-$p_{T}$ isolated leptons of different flavor, two b quark jets, and
    large missing $E_{T}$. We try to use variables which have this
information. The Monte Carlo samples consist of $\sim$1500 WW events
(background) and $\sim$3400 $t\bar{t}$ events (signal). 
The generation was done using CompHEP\cite{comphep} + 
Pythia hadronization\cite{pythia} + PGS\cite{pgs}. 
The following cuts were made to produce the samples used for
training and testing: $E_T^e > 15$ GeV, $E_T^\mu > 15$ GeV, missing $E_T > 15$
GeV, and two jets each with $E_T > 15$ GeV. The leptons and jets are
required to be centrally located.
Figure \ref{fig3label}a shows the four variables chosen because of
their potential discriminating power in a conventional cut-based
analysis. An exhaustive study of possible input variables was not done.
A training sample with these four variables was used to train the
SVM algorithm using the package LIBSVM \cite{libsvm} with a Gaussian
kernel function. The target
was 0.0 for the WW events and 1.0 for $t\bar{t}$ events. The SVM
output on an independent sample of events, shown in Figure
\ref{fig3label}b, peaks near zero for WW and closer to one for
$t\bar{t}$. The figure also
shows the performance of the SVM compared to 
conventional cuts. The SVM performance is about equal to the best
performance of the cut-based approach.

\begin{figure}
\begin{center}
\hspace{-1.5 in}
\begin{minipage}[t]{1.5in}
\includegraphics[width=7.6cm]{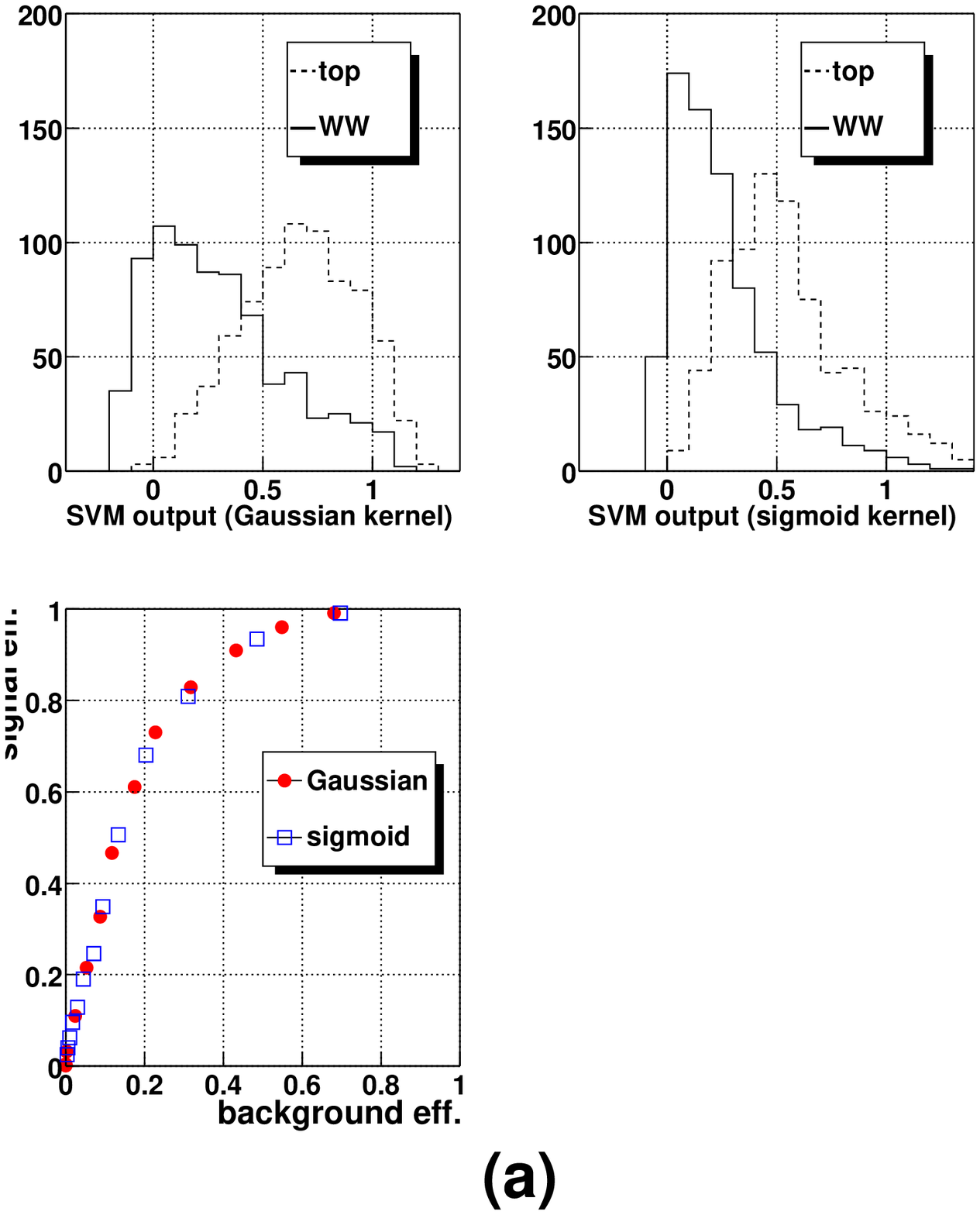}
\end{minipage}
\hspace{1.8 in}
\begin{minipage}[t]{1.5in}
\includegraphics[width=7.5cm]{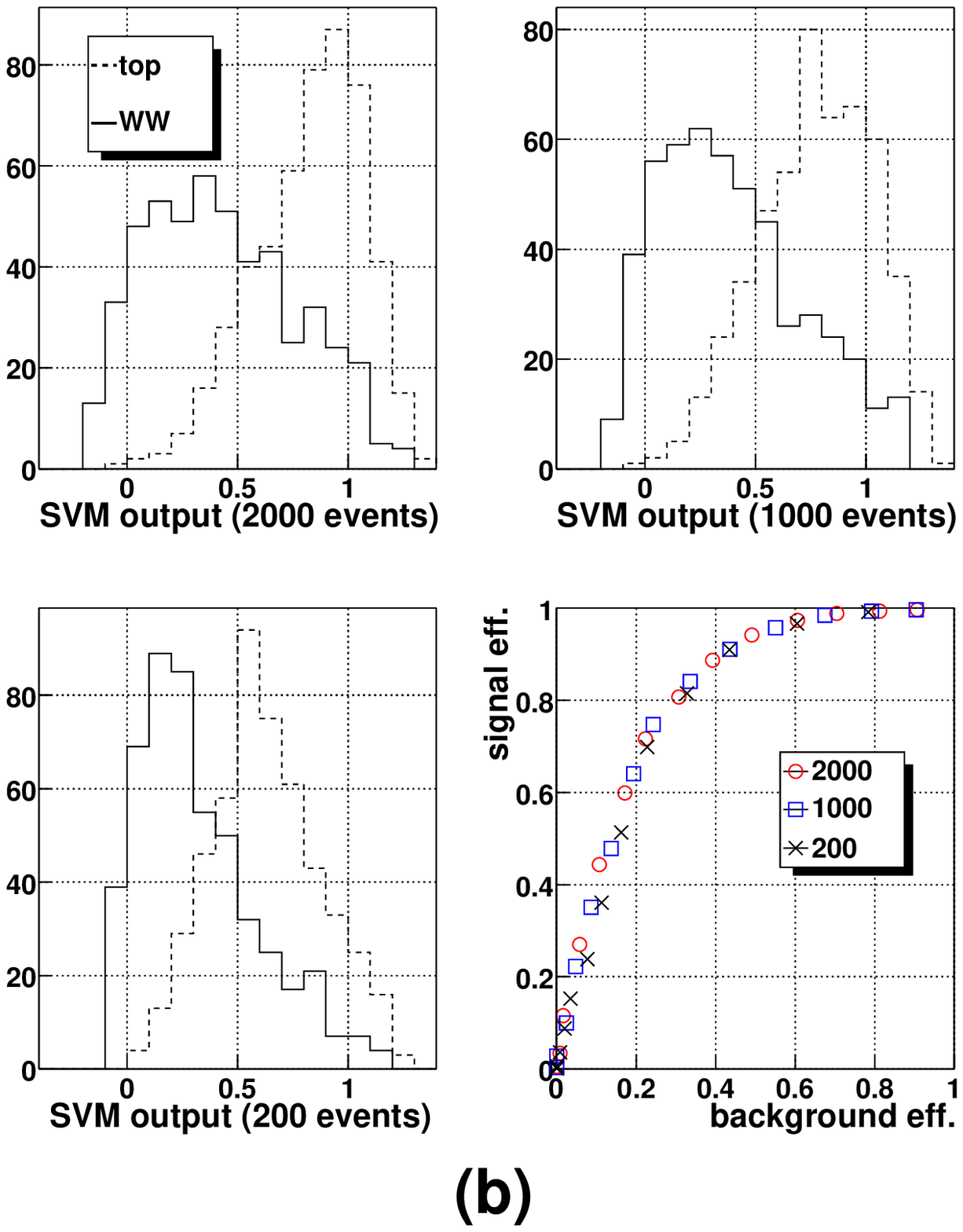}
\end{minipage}
\caption{(a) SVM performance of Gaussian kernel versus sigmoid kernel. 
(b) SVM performance for different numbers of training examples.}
\label{fig4label}
\end{center}
\end{figure}

Figure \ref{fig4label}a compares SVM performance with a Gaussian kernel 
and a sigmoid kernel. There is no significant difference.
Similarly, Figure \ref{fig4label}b shows that there is no significant
SVM performance difference as the number of the training examples varies
from 2000 to 200.

Training time must be considered for any supervised learning algorithm. A very 
long training time would make it difficult to thoroughly test the 
algorithm. The table below shows the training time in seconds for different
sample sizes. The right column shows the times for the $t\bar{t}$ + WW
Monte Carlo samples. The left column shows times for a toy Monte Carlo
sample with higher statistics.
The training was done with a gaussian kernel on a 1000 MHz Pentium III PC running
the Linux operating system. LIBSVM \cite{libsvm} uses a modified Sequential 
Minimal Optimization (SMO) algorithm that is known to be a fast method to 
train SVMs. The time scales roughly as the square of the number of
training examples. For the top quark study, training time was negligible.

\begin{center}
\begin{tabular}{c|cc}
number of events & toy MC time (sec) & $t\bar{t}$ MC time (sec)\\ \hline
200   &           $< 1$  &  $< 1$ \\
400   &           $< 1$  &  $< 1$ \\
1000  &          $< 1$   &  1 \\
2000  &             1.5  &  2 \\
3000  &             4    &  6 \\
4000  &             8    &  \\
8000  &            38    &  \\
16000 &           149    &  \\
\end{tabular}
\end{center}

\section{CONCLUSION}

SVMs provide nonlinear function approximations by mapping input vectors
into a high dimensional feature space where a hyperplane is constructed
to separate classes in the data. Computationally intensive calculations
in the feature space are avoided through the use of kernel functions.
SVMs correspond to a linear method in feature space which makes them
theoretically easy to analyze. The grounding in statistical learning 
theory leads to optimized generalization.
Advantages of SVMs include the existence of only one user chosen parameter
(aside from kernel parameters) and a unique, global minimum.
In an application of SVMs to top quark analysis we found that a 
straightforward application of the SVM algorithm
quickly reproduced the best performance of a cut-based approach.
SVMs are a new way to tackle complicated problems in high energy physics 
and other fields and should be considered as another technique for our 
multivariate analysis tool box.

\vskip1cm
\noindent

\section*{ACKNOWLEDGEMENTS}

I thank the organizers of the ``Advanced Statistical Techniques
in Particle Physics'' conference, the members of the Run II Advanced Algorithms
Group at Fermilab, and my colleagues in the CDF collaboration.

\end{document}